# *Processing Bias: Extending Sensory Drive to Include Efficacy and Efficiency in Information Processing*

*Julien P. Renoult*[1] *& Tamra C. Mendelson*[2]

[1] Centre of Evolutionary and Functional Ecology (CEFE UMR5175, CNRS—University of Montpellier—University Paul-Valery Montpellier—EPHE), 1919 route de Mende, 34293 Montpellier, France. email: julien.renoult@cefe.cnrs.fr

[2] Department of Biological Sciences, University of Maryland Baltimore County, 1000 Hilltop Circle, Baltimore, MD 21250, USA.

## Abstract

Communication signals often comprise an array of colors, lines, spots, notes or odors that are arranged in complex patterns, melodies or blends. Receiver perception is assumed to influence preference and thus the evolution of signal design, but evolutionary biologists still struggle to understand how perception, preference, and signal design are mechanistically linked. In parallel, the field of empirical aesthetics aims to understand why people like some designs more than others. The model of processing bias discussed here is rooted in empirical aesthetics, which posits that preferences are influenced by the emotional system as it monitors the dynamics of information processing, and that attractive signals have either effective designs that maximize information transmission, efficient designs that allow information processing at low metabolic cost, or both. We refer to the causal link between preference and the emotionally rewarding experience of effective and efficient information processing as the processing bias, and we apply it to the evolutionary model of sensory drive. A sensory drive model that incorporates processing bias hypothesizes a causal chain of relationships between the environment, perception, pleasure, preference, and ultimately the evolution of signal design, from simple to complex.

> **Glossary**
> **Efficacy:** optimizing information processing at any cost.
> **Efficiency**: information processing with an optimal use of resources.
> **Feature**: measurable property of a stimulus.
> **Information**: property of a stimulus that reduces uncertainty about the environment.
> **Information processing**: describes how information is received, transmitted in the nervous system, coded, stored and retrieved in the animal brain and sensory systems.
> **Neuronal selectivity**: the range of stimulus features that activate a neuron.
> **Processing bias**: judgment modulated by affect, which is influenced by the level of efficacy and efficiency in information processing. In cognitive sciences, processing bias is often referred to as an aesthetic judgment.
> **Sensory drive**: the hypothesis that the tuning of perceptual and cognitive systems to effectively and efficiently process information in environmental stimuli generates selection on communication signals due to a direct effect of effective and efficient processing on receiver preference.
> **Stimulus**: component of the external environment causing a physiological response (e.g., a landscape, an individual or a communication signal).

# 1. The complexity and diversity of communication signals

Evolutionary biologists continue to puzzle over the evolution of elaborate communication signals (Figure 1). Explanations are dominated by three hypotheses that are not necessarily mutually exclusive. One describes communication signals as quality indicators, whereby some feature (see Glossary) of the signal correlates with the fitness of the signaler (e.g., [1, 2]); for example, when the healthiest males express the most extreme sexual ornaments. Another derives from the verbal models of Fisher [3], in which signals become exaggerated and diversified simply as a result of genetic covariation with receiver preferences (e.g., [4, 5]). The third highlights the role of sensory perception and cognition, as in models of pre-existing bias and sensory drive [6-9]. Models of pre-existing bias assume that preferences evolve in a context other than mating, and communication signals that subsequently match those preferences are favored. Sensory drive emphasizes the importance of the environment in shaping perception and thus preferences. For example, if animal visual systems are tuned to local light conditions, the most effective visual signals will be those that maximally stimulate that particular tuning (e.g., [10]).

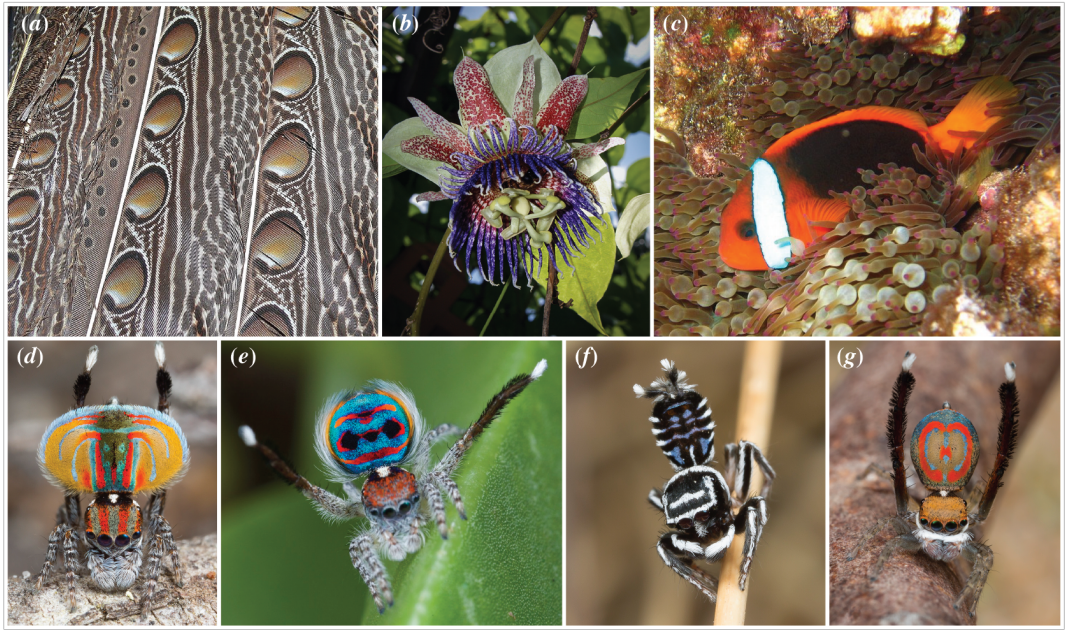

**Figure 1. Open questions about visual signals**. The design of visual communication signals has puzzled biologists since Darwin, who agreed with his contemporaries that feathers of the great argus pheasant ((*a*), *Argusianus argus,* credit: Bernard Dupont) were "more like a work of art than of nature" [11] (p258). The design of a signal—the arrangement of its features—can be complex, as in the flower of *Passiflora maliformis* ((*b*), credit: Nick Hobgood), or simple, as in the clownfish *Amphiprion melanopus*, ((*c*), credit: Richard Ling). Regardless, in most cases, there is no clear answer as to why the features of a signal are arranged as they are. Design diversification is just as enigmatic, as in the *Maratus* genus of jumping spider ((*d-g*); credits: Jurgen Otto). Why should the abdomen of one species feature red vertical bars surrounded by yellow patches transected by thin, curved blue lines (*d*), while its close congener sports dark eyespots and thick horizontal red bars on a background of turquoise blue surrounded by a yellow margin (*e*)? Extreme signal diversification is commonly attributed to the model of Lande [4], which hypothesizes that the trajectory of signal evolution changes with drift, or chance changes in female preferences; however, drift alone would ultimately produce random noise. It is generally accepted that receiver psychology influences the selection of signal design, and in some cases selected designs are more effective in transmitting information than non-preferred designs [12]. Why this occurs is still unknown. For two decades, cognitive scientists have studied the efficacy and the efficiency (see text) of artistic paintings in stimulating the perceptual, cognitive and emotional systems of humans. We propose that their results can shed light on the evolution of nature's designs.

Several authors have proposed that quality indicators and Fisherian models refer to the strategic component of signals, whereas perception models refer to signal efficacy [6, 12]. The strategic component is the actual content, the information being conveyed; efficacy describes the ability of a signal to reliably transmit the strategic component and thus refers to its form, or design. Here, we leverage a growing body of literature in the cognitive science of human aesthetics to argue that an additional component of signal design—efficiency, *the ability to convey information at low metabolic cost*—in addition to efficacy, is likely to play a major role in shaping signal design, from simple to complex (Figure 1). Both effective and efficient information processing can influence preference, via monitoring of information processing by the emotional system (Box 1). We refer to this incidental effect of information processing on preference as a "processing bias," and suggest that it can explain why effective and efficient signal designs are preferred. Unlike current bias models that focus primarily on efficacy and explain signal features that are easily detected, like color or contrast, a processing bias model accounts for efficiency and can explain some of the most enigmatic and complex signal patterns in nature.

## 2. Two fundamental aspects of information processing and signal design

Claude Shannon's information theory undoubtedly resides in the pantheon of scientific theories that have dramatically impacted civilization [13]. Information theory addresses two fundamental aspects of information processing: transmission and compression (reviewed in [14]). During transmission, random errors can be introduced in a signal, which make information noisy. *Efficacy* defines the ability of an information processing system to minimize noise and thus to maximize information transmission. Compression occurs because communication channels, i.e. the physical transmission medium, are often limited in their carrying capacity, or limited energetically. Compression is allowed by the presence of redundant information in a signal, and thus is often referred to as 'redundancy reduction'. The term *efficiency* is classically used to describe the ability of an information processing system to maximize information compression [14, 15].

Optimal information processing should simultaneously maximize efficacy and efficiency, and this dual maximization is at the heart of all modern information processing technologies. For example, many pixels in a digital visual image are redundant—they are the same as or easily predicted by the value of adjacent pixels. JPEG compression takes advantage of that redundancy; instead of reproducing the precise value of every pixel of a raw digital image, JPEG essentially smoothes values across adjacent pixels, transmitting one bit of information instead of many. But because an image still needs to be informative after decompression, the loss of information in a JPEG is barely detectable to a human eye. Importantly, however, maximizing efficacy and efficiency are two competing goals. Increasing compression (efficiency) adds noise, which degrades information transmission (efficacy). Consequently, there is a trade-off between efficacy and efficiency, and information processing has no absolute optimum. The optimal solution depends on the relative importance of either efficacy or efficiency relative to the goal pursued. JPEG processing software, for example, often permits users to manually set a compression level, depending on whether one wants to store a few heavy but high-quality, artistic pictures, or many vacation photos using minimum space.

Perceptual and cognitive systems appear to have been selected for optimal information processing. Selection for *efficacy* is supported by numerous adaptations that increase signal intensity (e.g., the summation of signals conveyed by multiple neurons; [16]) or decrease noise (e.g., the averaging of signals conveyed by multiple neurons; [17]). For example, in a variety of terrestrial and aquatic animals, photoreceptors are tuned to the lighting environment [10, 18, 19]. This tuning increases the signal-to-noise ratio, and thus the ability to detect or discriminate among stimuli. The tuning of photoreceptors to ambient light, and the adaptation of communication signals to maximize conspicuousness or detectability (i.e., efficacy), are some of the strongest evidence in support of sensory drive [20].

Selection for *efficiency* has been well documented in neuroscience. Attneave [21] and Barlow [22] were the first to apply the information theoretical definition of efficiency to animal perception, hypothesizing that animal brains reduce redundancies to provide an 'economical description' of the world. Information processing is a heavy metabolic cost: in humans, neuronal activity in the visual system alone accounts for 2.5 to 3.5% of a resting body's overall energy requirements [23]. Reducing the amount of neuronal activity required to process information should thus increase both efficiency and evolutionary fitness.

Brain adaptations to reduce redundancies have been studied mainly in visual communication, because visual stimuli naturally present a high level of spatial redundancy. Spatial redundancy can be characterized by lower- and higher-order statistics, both of which are processed in early stages of visual perception (in mammals: from the retina to the first visual cortical area [24-27]). Two examples of lower-order statistics are the spatial auto-correlation function and degree of scale-invariance; neurons in early stages of visual perception have adapted to these lower-order statistics of natural stimuli [28-33]. Higher-order statistics include sparseness, which is a measure of the neural activity required to encode a scene [33, 34]. Natural stimuli are particularly sparse [33], and visual modeling suggests that a critical function of early visual perception is to leverage this sparseness to efficiently process natural stimuli [33, 35].

## 3. Efficacy and efficiency influence preference

Research to date therefore indicates that perception and cognition utilize multiple strategies to both effectively and efficiently process information. In parallel, a growing body of literature on aesthetics demonstrates that effectively and efficiently processed stimuli are attractive to both humans and other animals. A link between information processing and preference has been documented in two subfields of aesthetics research: experimental psychology, which analyzes the effects of putatively aesthetic stimuli on behavior; and computational aesthetics, which addresses the spatial redundancy of aesthetic stimuli [36]. Over the last two decades, experimental psychologists have uncovered the 'fluency effect', by which people are attracted to stimuli that are fluently processed in the brain [37-41]. Fluency can be defined as the subjective experience of ease or difficulty in completing a mental task [42], and stimulus features associated with fluency suggest that it encompasses both efficacy and efficiency.

### (a) *Preference for effective stimuli*

The most well-known examples of preferred stimulus features are conspicuousness and symmetry, clearly associated with efficacy in information processing. Given the choice between conspicuous versus low-contrasting circles, people tend to prefer conspicuous circles [37]. Conspicuousness affects preference even when the design of the signal should play no role; for example, people are more trusting and more willing to follow instructions of a text, and find it more pleasant, when written in highly contrasting font [42]. Preference for conspicuous color stimuli occurs not only in humans, but also insects [43] and birds [44].

Symmetry is preferred in humans [45] and in many other animals [46]. In evolutionary biology, symmetry is often thought to be preferred because it indicates developmental quality [47]. However, symmetry also facilitates object detection and recognition and thus increases efficacy [46, 48]. For example, in a study of newly hatched chicks, naïve individuals innately preferred *asymmetric* geometric forms. Preference for symmetry appeared only in chicks that were allowed to forage independently; chicks that were hand-fed by researchers never developed a preference for symmetry [49]. This suggests that a key factor in symmetry preference was the improvement of sensorimotor skills during active food manipulation [49], rather than an innate preference typically assumed by indicator models. Thus the efficacy of symmetrical features itself likely influences preference, independently of the role of these features in signaling quality.

Another universal preference exists for prototypes, the most representative stimuli of a perceptual category. In addition to being effective (prototype-like stimuli are most quickly and precisely categorized and stored the longest in memory [50]), they are also the most attractive. Prototype preference has been shown in humans, using various biological, inanimate or abstract visual stimuli [50-52], and in other animals, notably in studies of the 'peak shift effect', when animals prefer an exaggerated version of the feature that distinguishes two perceptual categories [53]. For example, if a rat is trained to choose a rectangle with a 4:3 aspect ratio over a square, in subsequent testing trials the rat will choose a 3:2 rectangle over a 4:3 rectangle. The 3:2 rectangle is preferred because, in this example, it prototypifies rectangularity–the rule that the rat learned in order to differentiate a rectangle from a square–more than the 4:3 rectangle does [54].

In evolutionary biology, prototypes are often thought to be attractive because they exemplify features that define a fitness-related (i.e. quality) category; e.g., men should prefer the most feminine women because femininity indicates fertility [55]. However, a study in chickens analyzed preference for human faces: birds that were trained to choose the average female face from a range of female and male faces were found to respond maximally to female faces that were more feminine than average during testing trials, and specifically to the face that was also rated as being most beautiful by human subjects [56]. This and other results (e.g., [50]) suggest that prototypes are preferred in part because they increase processing efficacy independently of the quality of the stimulus.

Repetition over time is another feature that increases preference. Repeated stimuli are effective because they provide prior knowledge about the structure of the signal, which allows neurons to anticipate and compensate for noise [17]. Known in psychology as the 'mere exposure effect' [57], people tend to prefer repeated stimuli over stimuli to which they have never been exposed. For example, the mere repetition of a melody is sufficient to increase preference for it, at least in initial stages [58].

(b) *Preference for efficient stimuli*
Psychological studies on the fluency effect also show that preferred stimuli have features promoting efficiency. Prototypicality and repetition over time are efficiently processed in addition to being effective. Prototypes are sparsely encoded and thus economical because they only need to stimulate a few highly selective neurons to be recognized [59]. Repetition also increases sparseness because the selectivity of neurons tends to be tuned to features to which they are frequently exposed [60]. Surface and line continuity also promote efficiency. Continuous surfaces/lines are redundant and thus highly predictable, and are preferred over discontinuous lines and heterogeneous surfaces [61]. Preference for continuous shapes also has been shown in non-human primates [62] and birds [63].

Computational aesthetics provides more direct evidence that sparseness elicits preference. A common way to measure the sparseness of an image is to first train artificial neurons to process images of natural scenes while minimizing the number of simultaneously active neurons (a sparseness constraint) [33, 35]. This creates a proxy network for the first visual cortex of mammals (V1) that is 'adapted' to sparsely encode terrestrial environments, which is then used to estimate the sparseness of images: sparse images will activate fewer neurons in the trained network. Renoult et al. [64] used this approach to model the sparseness of images of female faces. Sparseness was positively correlated with face attractiveness as rated by men and explained up to 17% of the variance in attractiveness. Using the same approach, Holzleitner et al. [65] found that sparseness was the highest predictor of face attractiveness when compared to body mass index, sexual dimorphism, averageness and asymmetry. It is worth noting that features unrelated to efficiency could drive sparse coding in V1 (e.g., a smooth skin texture that indicates youth and health); these results therefore would benefit from analyzing sparseness in higher levels of information processing. Nevertheless, in another study that directly modeled V1 from neurophysiological data, image sparseness was negatively correlated with aversiveness: images of abstract patterns with a lower degree of sparseness were more highly aversive to human subjects [66]. Thus, images that are sparsely processed by environmentally tuned visual systems (i.e., trained neural networks) are attractive.

Additional support for a link between efficiency and preference comes from computational studies of artwork. Natural terrestrial scenes are characterized by an elevated and characteristic degree of scale-invariance, often measured as the fractal dimension *D* or as the slope of the Fourier power spectrum 1/f [14, 50, 67]. Several studies demonstrate that people prefer both abstract and representational images with fractal dimensions that mimic those of an average natural terrestrial scene [32, 68, 69]. Other studies demonstrate that artistic paintings have a degree of scale-invariance similar to that of natural scenes; for example, faces painted by portrait artists across time and cultures exhibit natural values of scale invariance even though real faces typically do not [70]. It has been suggested that artists increase the attractiveness of their work by unconsciously mimicking the spatial statistics of natural scenes to which our brain has adapted to efficiently process, thus creating art that is 'easy-on-the-eyes' [31, 71, 72].

## 4. The processing bias

Cognitive research therefore strongly suggests that people and other animals prefer effective and efficient visual stimuli, and moreover that effective and efficient processing is a pleasant experience [73]. One explanation for this observation, rooted in the quality indicator model of sexual selection, is if neurons are tuned to stimuli with the greatest impact on receivers' fitness. This explanation is related to the psychological hypothesis that perceptual and cognitive fluency may be pleasant because it indicates that a stimulus is familiar and thus less likely to cause harm [57, 74]. In this case, effective and efficient processing would indicate the value, or quality, of the signaler; signalers of the highest quality will display the most optimally processed traits. Receivers that are rewarded for attending to those traits are assumed to prefer them and seek them out; thus, selection should favor a pleasurable response to effective and efficient processing. A quality indicator explanation for pleasure in information processing therefore implicates the stimulus as the selective agent shaping perceptual systems, rather than vice versa, or at least playing a lead role in a co-evolutionary process. It furthermore presupposes that a receiver evaluates the stimulus positively, and responds to it appetitively, thus establishing a link between pleasure and preference. Importantly, however, not all stimuli that affect receiver fitness trigger a positive emotional evaluation. Aposematic traits, for example, facilitate the recognition and memorization of dangerous species [75] and are judged to be beautiful, but they trigger fear, a negative or aversive emotional evaluation (e.g., in snakes, see [76]; Box 1). Animals can form positive (aesthetic) judgments about stimuli that trigger an aversive emotional response or even about stimuli that are likely neutral with respect to fitness, like artwork or landscapes, suggesting a typical quality indicator model may not be a sufficient explanation.

An alternative explanation refers to the 'meta-informative function' of pleasure, which helps monitor progress in information processing and is triggered when processing is effective and efficient ([73, 77]; Box 1). A meta-informative function of pleasure is supported by a corpus of studies showing that information seeking and problem solving are experienced as intrinsically pleasurable activities ([78]; see also [79]). This pleasure is likely adaptive—effective and efficient processing can yield direct benefits like reducing response time and decreasing metabolic costs. Unlike a standard quality indicator hypothesis, however, the meta-informative function of pleasure does not necessarily associate pleasure with desire, which is consistent with psychological research in liking and wanting. Though we generally "like what we want, and want what we like" [80], pleasure and desire are mediated by different neurotransmitters that can be released separately [81]. Cognitive scientists have investigated whether and why people (or other animals) might find stimuli pleasant solely due to information processing itself and not from an evaluation of stimulus benefits. The answer appears to be misattribution, which describes the fact that people are usually unaware of the source of pleasure and, by default, tend to attribute it to the stimulus rather than to information processing itself [42, 50]. Misattribution of pleasure therefore implies that the link between efficient processing and preference is a sensory or cognitive 'bias', that is, *preference is a by-product of the adaptive function of pleasure to monitor progress in information processing.* We call this the *processing bias.*

Like other pre-existing biases, processing bias implicates perceptual systems as the driver of signal evolution, rather than vice versa, specifically as signals evolve to leverage or exploit a pre-existing bias for effective and efficient information processing. That bias will be shaped by the environment not only to maximize signal detection and discrimination but also to minimize habitat-specific redundancies, which can be exploited by the patterns and textures of animal signals. Because it is distinct from emotional or cognitive evaluation, processing bias can be demonstrated by, if not a preference, at least a positive attitude (e.g, curiosity) toward effective and efficient features that persists after uncoupling efficacy and efficiency with rewards, or by showing that reversing a preference through associative learning is more difficult when stimuli are effectively and efficiently processed (see also [82]).

**Box 1. How animals process information**
Information processing describes the mechanisms that produce a behavioral output from a stimulus input. For most behavioral outputs, animals do not simply reflexively respond to external events, or stimuli; rather, they build meaning by extracting and transforming information from these stimuli. Information processing requires three brain systems: perception, cognition and emotion (**Figure B.1,** [80, 83]).

Perception is the foundational system of information processing and its function is to build an internal representation of the external world. This is achieved by first converting a stimulus into a neural code, and then by hierarchically extracting information from this code. The extracted information is increasingly complex (e.g., simple line segments in early visual stages and entire objects in higher stages) and global (e.g., neurons respond to stimuli spanning the whole visual field only in higher stages; [84]). Cognition is the brain system where highly integrated processes occur. It helps build a meaningful representation from perception by providing knowledge about the environment, which notably requires memory. Information processed by perception and then by cognition gives rise to a cognitive evaluation of a stimulus (along a continuum of negative to positive) that indicates the costs or benefits of the stimulus for the receiver.

The third brain system, emotion, also gives rise to an evaluation, consciously experienced or not, along a continuum of negative to positive, reflecting the receiver's interaction with the environment [85]. For example, fear of predators is a negative emotional evaluation that reflects a highly costly interaction. Like its cognitive counterpart, the emotional evaluation influences preference, and the behavior [86]. The emotional and cognitive evaluations have nevertheless distinct neurochemical bases, and most importantly they differ in the timing of their effects, the emotional evaluation developing earlier during information processing than the cognitive evaluation [87].

Emotions are determined by affects, which play an important role in informing the receiver about the rate of progress toward a goal, and reward it for successful progress [77]. The core rewarding affect is pleasure [88]. In addition to mediating the emotional evaluation, affects also have a meta-informative function: they evaluate progress in information processing [77] and thereby help regulate the process of information gathering. Depending on how pleasurable information processing is, the receiver will continue the same processing strategy, change its strategy, or stop processing information [77].

The tri-partite model of information processing is a highly simplified description of how animal brains process information. Yet it has two main advantages that make it useful for evolutionary biology. First, it excludes brain processes that are still hotly debated among cognitive scientists, such as the relative importance of feedback interactions between cognition and perception [89]. Second, the model likely applies to most if not all brained animals. Even tiny brains such as those of insects are capable of complex cognitive operations (reviewed in [90]) and emotions. Compared to cognition, non-human emotions have been historically more controversial, but interest in their study has increased in recent years, with the development of experimental frameworks for their analysis [85, 91]. For example, using an experimental approach similar to those used in humans to study pessimism and optimism (an 'half-full vs. half-empty glass' approach), a recent study found that bees who experienced a punishing or a rewarding event were more likely to subsequently respond negatively or positively, respectively, to an ambiguous task [92]. As in humans, these animal emotions are modulated by affects [92], which also monitor the dynamics of information processing [93].

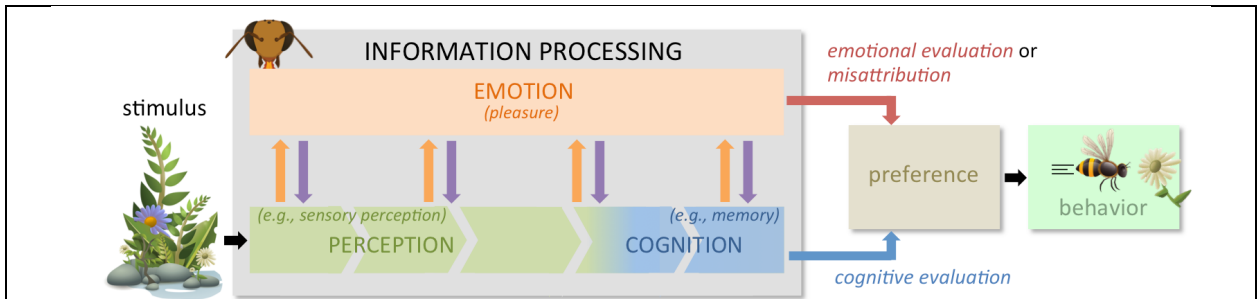

**Figure B.1. Information processing in animal brains**. The information conveyed by a stimulus (e.g., a flower) is processed by perceptual and then cognitive neurons of the receiver (e.g., a bee), leading to a cognitive evaluation of the costs and benefits of the interaction outcome (e.g., quantity of nectar; blue arrow). Along the processing pathway, pleasure is triggered when processing is effective or efficient (e.g., conspicuous flower; orange arrows). This pleasure could contribute to a fast emotional evaluation of the costs and benefits of interacting with the signaler or of the direct energetic benefits of processing an efficient stimulus (red arrow). Alternatively or in addition, pleasure can result from evaluating progress in information processing and thereby help regulate the process of information gathering [77](violet arrows). Because the receiver is not aware that pleasure is triggered by efficient processing, by default s/he misattributes it to the stimulus, which may bias preference toward this stimulus (red arrow).

**Box 2. Evolutionary aesthetics**

Unraveling the functional bases of aesthetics has been a major research aim in cognitive sciences over the last two decades. Although no unique definition of aesthetics has yet emerged, results overall agree that aesthetic experiences are affective, independent of the sensory modality and the perceptual domain, and rooted in the interaction between a stimulus and a receiver (i.e., any object, organism or even landscape could operate aesthetically [94, 95]). The German philosopher Immanuel Kant and others further suggested that the aesthetic experience is a state of "disinterested interest", an engagement with objects without the desire to acquire, control or manipulate them [80]. This idea has been supported recently by brain imaging studies (e.g., [96]).

A large body of work suggests that the pleasure experienced by efficient information processing fulfills all criteria of an aesthetic experience. Studies show that artists select, consciously or not, formal (e.g., colors and patterning) or conceptual features to manipulate the efficiency of perceptual or cognitive processing, respectively [31, 68-70]. Artists also develop complex strategies to delay the acquisition of information (e.g., suspense), or initially confuse the receiver in order to amplify the aesthetic pleasure of a sudden increase in processing efficiency (the 'Aha' effect [74]).

The processing bias model discussed here is an evolutionarily-framed extension of the models of fluency [40, 45], efficient coding [42] and pleasure-interest model of aesthetic liking [41] proposed by cognitive scientists to explain the results described above. The popularity of these models has grown among biologists, social scientists and philosophers due to their unique ability to simultaneously account for both the universal and subjective dimensions of aesthetics, and to explain complex aesthetic experience beyond controlled, laboratory conditions [73].

> Extending the scope of aesthetics from the arts to natural communication, some authors have proposed that many signals in plants and animals could be aesthetic [75-77]. For example, the peacock tail trapping peahens' gaze into a back-and-forth motion [78], or the visual illusions of bowerbirds [79] could be strategies to manipulate pleasure caused by efficient information processing. As a cautionary note, however, we stress that the similarity between artwork and natural aesthetic signals is homologous in process (i.e., they have similar biological bases), not in function [77]. The processing bias model thus does not predict that works of art function as sexual signals.
>
> One aspect of aesthetic communication in plants and animals that remains to be determined is whether receivers actively search for an aesthetic experience. The human parallels are museum visitors, moviegoers and book readers who spend time and money to be rewarded by nothing more than the pure pleasure of an information processing experience [77]. In other words, mates or pollinators may select partners or plants for their aesthetic reward in addition to the benefits of these resources.

## 5. Importance of efficiency for the evolution of communication
(a) *Extending sensory drive to efficiency*

Sensory drive describes the influence of the external environment on the design of communication signals through its effects on perception and cognition [6, 9]. To date, sensory drive has been framed primarily to explain the efficacy of signals, where signal evolution is driven by neurons that are tuned to maximize detection, discrimination or recognition in a particular set of environmental conditions. Here, we extend the model to include efficiency, because neurons are also tuned to efficiently process the characteristic redundant features of their habitats. Just as artists mimic spatial features of natural scenes to make their artwork more attractive, a sensory drive model that incorporates efficiency predicts that organisms have evolved communication signals that match the lower- (e.g., the degree of scale-invariance) and higher-order (e.g., sparseness) statistics of their environments.

Recognizing the importance of efficiency in signal design will likely increase our estimate of the role of sensory drive in evolution. In visual communication, canonical examples of sensory drive come from aquatic habitats, which vary in the color of ambient light (e.g., [10, 97]). In contrast, terrestrial habitats vary little in ambient light, such that the role of sensory drive in terrestrial species has remained contentious (e.g., [98]). Studying spatially redundant features could provide broader support for a role of sensory drive in signal evolution, because these features differ strongly across both terrestrial [30] and aquatic habitats [Hulse et al., in prep].

Historically, sensory drive has focused on how signal detection (efficacy) is shaped by the transmission channel, for example the color of ambient light or the acoustic characteristics of background noise. Perception and cognition can adapt to other aspects of the environment as well, however, for example food items [18] or sexual displays [99]. Neuronal tuning thus likely reflects adaptation to many biotic and abiotic environmental stimuli, creating multiple efficacy 'niches' to which signals can adapt (i.e., exploit), e.g., by evolving conspicuous colors, symmetrical and prototypical patterns or combinations of these features [8]. The same reasoning holds for efficiency. In the visual system, for example, neurons in the eyes are tuned to simple, redundant features of habitats (i.e., simple oriented line segments); however, later in the processing pathway neurons are tuned to efficiently process more complex and specific features (e.g., familiar faces; see Box 1). Multiple stages of information processing thus also create multiple efficiency 'niches' to which signals can adapt, as local environments vary not only in spatial statistics but also community composition (which can affect the features used to efficiently classify categories like mate, competitor, or predator). A sensory drive framework that incorporates both efficacy and efficiency therefore provides an even more powerful explanation for the diversification of animal signals.

(b) *Studying efficiency in evolution*

A variety of empirical approaches can test whether communication signals have evolved to be efficiently processed, and furthermore, whether the adaptation of neural systems to efficiently process local habitats drives the evolution of signal design (i.e., sensory drive; Box 3). These approaches will mirror those that test the efficacy component of sensory drive, which focus mainly on signal detection and discrimination. Assuming that efficiency-driven preference originates from redundant features of the habitat, the sensory drive model predicts that redundant features of signals should match redundant features of habitats. The model thus predicts interspecific variation in patterns and texture will be correlated with variation in habitat, and convergence will occur between unrelated species living in similar habitats.

A sensory drive model of efficiency can also be tested intraspecifically. Here, predictions of sensory drive might differ from that of quality indicator models. Whereas indicator models predict that sexual signals will exhibit the most regular patterns (e.g., the highest fractal dimension), sensory drive based on a processing bias predicts that signals should evolve toward environmental-like statistics (e.g., natural values of auto-correlation, scale invariance, sparseness) rather than toward maxima. For example, the black bib of the red-legged partridge (*Alectoris rufa*) is a male sexual ornament whose spatial statistics (fractal dimension $D$) are associated with higher individual condition [67, 100], consistent with predictions of indicator models. But whether the fractal dimension of the bib closely matches that of its habitat, or whether female preference is predicted by the fractal dimension, remains to be tested.

The strongest support for sensory drive is a complete sequence whereby habitat features match neuronal tuning, neuronal tuning is correlated with preference, and preference is correlated with signal design. Such a sequence appears to have been established for efficacy in the African cichlid fish genus *Pundamilia*, where sister species live in blue- or redshifted light. *P. nyererei* lives in redshifted light, which increases the signal-to-noise ratio (conspicuousness) of red stimuli against a dark background. *P. nyererei* exhibit higher expression of long wavelength (red) sensitive photoreceptors [101] and are more sensitive to red light [102]. Female *P. nyererei* prefer red stimuli, and males have evolved reddish coloration [10]. As for efficiency, each of these correlations has been shown in one system or another (see above), but to our knowledge they remain to be shown altogether in a single system.

Camouflage patterns pose an interesting problem with respect to efficacy and efficiency. Because they match the spatial features of environments and are thus efficiently processed, camouflage patterns should be attractive to predators; yet, such designs are selected to be undetectable. Camouflage patterns are therefore at the same time efficient but almost completely ineffective (from an information processing perspective), so the pleasure triggered by efficient processing of camouflage is unlikely to have played a role in the evolution of the pattern or of the predator's decision. However, if predation pressure is relaxed or lost, for example as observed during island colonization, signalers are free to evolve conspicuous socio-sexual color signals on top of their ancestral camouflage. Pleasure derived from processing efficient camouflage patterns would then compound the pleasure of effectively processing conspicuous features, and signalers that have retained their ancestral camouflage should be preferred. Camouflage patterns could thus easily be co-opted for sexual signaling. Although a link between camouflage and sexual signaling has been ignored in the recent literature, the idea is not new. Renowned 19th-century artist and naturalist Abbott Thayer was once mocked for suggesting that all animal patterns, even the peacock's tail, are cryptic [103]. Investigating the links between efficacy and efficiency, and between sexual selection and camouflage, could partially vindicate that perspective.

**Box 3. Estimating processing efficiency in visual communication**
Efficiency characterizes information processing at low metabolic cost. Empirically estimating efficiency thus requires measuring the energetic cost of processing and comparing it between alternative processing strategies [104], or to the same strategy applied to structurally different but functionally similar stimuli (e.g., the sexual signals of different males in a population). In lab studies with primates and rodents, the standard approach is to analyze functional connectivity using brain imaging, which estimates whether the distance travelled by information throughout different brain areas is minimized [105]. The study of brain functional connectivity is limited to model species, however, and thus most studies in evolutionary biology would rely on more indirect methods.

Efficiency can be estimated indirectly with statistics that describe spatial redundancy in stimuli. The most well-studied and commonly used statistics are spatial auto-correlation and scale invariance, which can be estimated using Principal Component Analysis (PCA; [106]) for the former, and the 1/f spectral slope [36, 72] or fractal dimension $D$ [32, 67] for the latter. These statistics indicate the efficiency of information processing because animal perceptual systems have evolved to reduce spatial redundancies occurring in natural environments. Thus the most efficiently processed stimuli have spatial statistics matching most closely those of natural environments.

Processing efficiency also can be estimated using models of perception and cognition. Neurons selective to locally oriented line segments (as found, for example, in the primary visual cortex of mammals or in the tecto-isthmic area in fishes) can be computationally modeled using simple Gabor filters [107], or by training a set of basis functions (each one modeling one neuron) to encode images of visual stimuli as sparsely as possible [33]. Then, efficiency is modeled by estimating the sparseness of the neuronal responses to a stimulus image [26, 64, 71, 108]. Here, sparseness is measured as the proportion of neurons activated (i.e., with a non-zero response), or the kurtosis of the response distribution [109]. One limitation to this approach, however, is that efficiency is estimated at one level of neural processing only.

Convolutional Neural Networks (ConvNets) –the tool of choice for deep learning and artificial intelligence– are a promising approach for estimating efficiency throughout the processing pathway. Although the primary goal of ConvNets is not to reproduce the mechanisms behind animal perception and cognition, the different layers of a ConvNet have been found to accurately model multiple levels of neuronal processing [110]. ConvNets could thus be used to compare efficiency across early perceptual and higher cognitive processing by calculating the sparseness of neuronal activation at each layer of the network. Finally, computer scientists have recently used information theory to study the efficacy of information transmission across ConvNets [111]. By simultaneously estimating the efficacy and efficiency of processing a given stimulus, future research should be able to address how these two components interact to influence preference and the evolution of signal designs.

## 6. A broader outlook

Historically in evolutionary biology, the mechanisms that link stimulus and behavior have been modeled using simplified frameworks that ignore the complexity of brain processes [112]. This simplified approach has been motivated possibly by a fear of anthropomorphism, but also because brain processes were once considered elusive and unpredictable (but see [113, 114]). Advances in comparative cognition [115, 116], however, are allowing evolutionary biologists to investigate the ubiquitous role of cognitive processes in social, sexual and natural selection, and in speciation [113, 117-119]. We suggest that further accounting for emotions, and for efficiency in perceptual and cognitive processes, can address persistent outstanding questions in animal communication and beyond. In humans, for example, pleasure mediated by effective and efficient processing is also known to bias judgment of truth (review in [120]); e.g., reading an identical email three times rather than once increases estimates of how many people would agree with its content [121]. The consequences of a link between efficacy (and efficiency) and truth on the evolution of honesty, and more generally on the strategic component of communication signals, is an open field of study.

    A central point is that behaviors can be motived by rewards arising from the evaluation of neural processes independent of the evaluation of benefits provided by a signaler. Here, we examined the link between the environment, information processing, pleasure and preferences. Other types of intrinsically rewarding processes include curiosity, an adaptive behavior aimed at filling a gap in knowledge, for which reward also originates from information seeking itself and not from the good or bad use of that information [122]. Psychologists and philosophers have suggested that such rewards could be major determinants of aesthetic preferences and thus could help explain the diversity of works of art [123]. Studying intrinsically rewarding processes could similarly reveal how elaborate communication signals evolve in other species (Figure 1). Models of sensory drive and pre-existing bias offer especially appropriate evolutionary frameworks for integrating hypotheses and results from the humanities, cognitive psychology, neurophysiology, computer science and behavioral ecology to understand the evolution of signal design.

### Authors' contributions
The two authors contributed equally to this review.

## Acknowledgments
The authors thank M. Burns, S. Hulse, B. Lohr, K. Omland and other members of the UMBC ecology and evolution journal club, B. Godelle, D. Gomez and other members of the CEFE and ISEM journal clubs in Montpellier, other friendly reviewers.

## Funding
The collaboration between authors is funded by National Science Foundation grant IOS-1708543.

## References

1. Andersson M. 1994 *Sexual selection*. Princeton, New Jersey, Princeton University Press; 599 pp.
2. Kokko H. 2001 Fisherian and "good genes" benefits of mate choice: how (not) to distinguish between them. *Ecoly Letters* **4**, 322-326.
3. Fisher R.A. 1930 *The Genetical Theory of Natural Selection*, Oxford University Press, Oxford.
4. Lande R. 1981 Models of speciation by sexual selection on polygenic traits. *Proceedings of the National Academy of Sciences* of the USA **78**, 3721-3725.
5. Prum R.O. 2010 The Land-Kirkpatrick mechanism is the null model of evolution by intersexual selection: implications for meaning, honesty, and design in intersexual signals. *Evolution* **64**, 3085-3100.
6. Endler J.A., Basolo A.L. 1998 Sensory ecology, receiver biases and sexual selection. *Trends in Ecology and Evolution* **13**, 415-420.
7. Schaefer H.M., Ruxton G. 2015 Signal diversity, sexual selection, and speciation. *Annual Review of Ecology and Evolution and Systematics* **46**, 573-592.
8. Ryan M.J., Cummings M.E. 2013 Perceptual biases and mate choice. *Annual Review of Ecology and Evolution* **44**, 437-459.
9. Endler J.A. 1993 Some general comments on the evolution and design of animal communication systems. *Philosophical Transactions of the Royal Society of London Biology B* **340**, 215-225.
10. Seehausen O., Terai Y., Magalhaes I.S., Carleton K.L., Mrosso H.D.J., Miyagi R., van der Sluijs I., Schneider M.V., Maan M.E., Tachida H., et al. 2008 Speciation through sensory drive in cichlid fish. *Nature* **455**, 620-626.
11. Darwin C. 1871 *The Descent of Man, and Selection in Relation to Sex*. London, John Murray.
12. Guilford T., Dawkins M.S. 1991 Receiver psychology and the evolution of animal signals. *Animal Behavior* **42**, 1-14.
13. Shannon C.E. 1948 A Mathematical Theory of Communication. *Bell System Technical Journal* **27**, 379-423.
14. Hyvärinen A., Hurri J., Hoyer P.O. 2009 *Naural Image Statistics: A Probability Approach to Early Computational Vision*. London, Springer-Verlag.
15. MacKay D.J.C., Mac Kay D.J.C. 2003 *Information theory, inference and learning algorithms*, Cambridge university press.
16. Stöckl A.L., O'Carroll D.C., Warrant E.J. 2016 Neural summation in the hawkmoth visual system extends the limits of vision in dim light. *Current Biology* **26**, 821-826.
17. Faisal A.A., Selen L.P.J., Wolpert D.M. 2008 Noise in the nervous system. *Nature Reviews Neuroscience* **9**, 292-303.
18. Osorio D., Vorobyev M. 1996 Colour vision as an adaptation to frugivory in primates. *Proceedings of the Royal Society of London Biology B* **263**, 593-599.
19. Osorio D., Vorobyev M. 2005 Photoreceptor spectral sensitivities in terrestrial animals: adaptations for luminance and colour vision. *Proceedings of the Royal Societyof London Biology B* **272**, 1745-1752.
20. Fuller R.C., Endler J.A. 2018 A perspective on sensory drive. *Current Zoology* **64**, 465-470.
21. Attneave F. 1954 Some informational aspects of visual perception. *Psychological Review* **61**, 183.

22. Barlow H. 1961 Possible principles underlying the transformations of sensory messages. In *Sensory Communication* (ed. Rosenblith W.), MIT Press, Cambridge.
23. Attwell D., Laughlin S.B. 2001 An energy budget for signaling in the grey matter of the brain. *Journal of Cerebral Blood Flow & Metabolism* **21**, 1133-1145.
24. Dan Y., Atick J.J., Reid R.C. 1996 Efficient coding of natural scenes in the lateral geniculate nucleus: experimental test of a computational theory. *The Journal of Neuroscience* **16**, 3351-3362.
25. Pitkow X., Meister M. 2012 Decorrelation and efficient coding by retinal ganglion cells. *Nature Neuroscience* **15**, 628-635.
26. Simoncelli E.P., Olshausen B.A. 2001 Natural image statistics and neural representation. *Annual Review of Neuroscience* **24**, 1193-1216.
27. Simoncelli E.P. 2003 Vision and the statistics of the visual environment. *Current Opinion in Neurobiology* **13**, 144-149.
28. Srinivasan M.V., Laughlin S.B., Dubs A. 1982 Predictive coding: a fresh view of inhibition in the retina. *Proceedings of the Royal Society of London Biology B* **216**, 427-459.
29. Atick J.J. 1992 Could information theory provide an ecological theory of sensory processing? *Network: Computational Neural Systems* **3**, 213-251.
30. Pouli T., Reinhard E., Cunningham D.W. 2014 *Image Statistics in Visual Computing*. Boca Raton, FL, Taylor & Francis Group; 354 p.
31. Taylor R., Spehar B., Hagerhall C., Van Donkelaar P. 2011 Perceptual and physiological responses to Jackson Pollock's fractals. *Frontiers in Human Neuroscience* **5**, 60.
32. Taylor R.P., Spehar B. 2016 Fractal fluency: an intimate relationship between the brain and processing of fractal stimuli. In *The Fractal Geometry of the Brain* (pp. 485-496), Springer.
33. Olshausen B.A., Field D. 1996 Emergence of simple-cell receptive field properties by learning a sparse code for natural images. *Nature* **381**, 607-609.
34. Field D.J. 1987 Relations between the statistics of natural images and the response proporties of cortical cells. *Journal of the Optical Society of America A* **4**, 2379-2394.
35. Olshausen B.A., Field D.J. 1997 Sparse coding with an overcomplete basis set: A strategy employed by V1? *Vision Research* **37**, 3311-3325.
36. Brachmann A., Redies C. 2017 Computational and experimental approaches to visual aesthetics. *Frontiers in Computational Neuroscience* **11**, 102.
37. Reber R., Winkielman P., Schwarz N. 1998 Effects of perceptual fluency on affective judgments. *Psychological Sciences* **9**(1), 45-48.
38. Reber R. 2012 Processing fluency, Aesthetic pleasure, and culturally shared taste. In *Aesthetic Science, Connecting Minds, Brains, and Experience* (eds. Shimamura A.P., Palmer S.E.), Oxford University Press, NY.
39. Winkielman P., Schwarz N., Fazendeiro T., Reber R. 2003 The hedonic marking of processing fluency: Implications for evaluative judgment. In *The sychology of evaluation: Affective processes in cognition and emotion* (eds. Musch J., Klauer K.C.), pp. 189-217, Psychology Press.
40. Graf L.K.M., Landwehr J.R. 2015 A dual-process perspective on fluency-based aesthetics: the pleasure-interest model of aesthetic liking. *Personality and Social Psychology Review* **19**, 395-410.
41. Redies C. 2007 A universal model of esthetic perception based on the sensory coding of natural stimuli. *Spatial Vision* **21**, 97-117.
42. Oppenheimer D.M. 2008 The secret life of fluency. *Trends in Cognitive Sciences* **12**, 237-241.
43. Naug D., Arathi H.S. 2007 Receiver bias for exaggerated signals in honeybees and its implications for the evolution of floral displays. *Biology Letters* **3**, 635-637.
44. Cazetta E., Schaefer H.M., Galetti M. 2009 Why are fruits colorful? The relative importance of achromatic and chromatic contrasts for detection by birds. *Evolution & Ecology* **23**, 233-244.
45. Reber R., Schwarz N. 2006 Perceptual fluency, preference, and evolution. *Polish Psychological Bulletin* **1**, 16-22.
46. Enquist M., Arak A. 1994 Symmetry, beauty and evolution. *Nature* **372**, 169-172.
47. Grammer K., Thornhill R. 1994 Human (*Homo sapiens*) facial attractiveness and sexual selection: the role of symmetry and averageness. *Journal of Comparative Psychology* **108**, 233-242.
48. Garner W.R. 1974 *The processing of information structure*. Potomac, MD, Lawrence Erlbaum Associates, Inc.
49. Clara E., Regolin L., Vallortigara G. 2007 Preference for symmetry is experience dependent in newborn chicks (*Gallus gallus*). *Journal of Experimental Psychology* **33**, 12-20.

50. Winkielman P., Halberstadt J., Fazendeiro T., Catty S. 2006 Prototypes are attractive because they are easy on the mind. *Psychological Sciences* **17**, 799-806.
51. Farkas A. 2002 Prototypicality-effect in surrealist paintings. *Empirical Studies of the Arts* **20**, 127-136.
52. Whitfield T.W.A., Slatter P.E. 1979 The effects of categorization and prototypicality on aesthetic choice in a furniture selection task. *British Journal of Psychology* **70**, 65-75.
53. ten Cate C., Rowe C. 2007 Biases in signal evolution: learning makes a difference. *Trends in Ecology and Evolution* **22**, 380-387.
54. Ramachandran V.S., Hirstein W. 1999 The science of art: A neurological theory of aesthetic experience. *Journal of Consciousness Studies,* **6**, 15-51.
55. Perrett D.I., Lee K.J., Penton-Voak I., Rowland D., Yoshikawa S., Burt D.M., Henzi S.P., Castles D.L., Akamatsu S. 1998 Effects of sexual dimorphism on facial attractiveness. *Nature* **394**, 884.
56. Ghirlanda S., Jansson L., Enquist M. 2002 Chickens prefer beautiful humans. *Human Nature* **13**, 383-389.
57. Zajonc R.B. 1968 Attitudinal effects of mere exposure. *Journal of Personality and Social Psychology* **9**, 1-27.
58. Pereira C.S., Teixeira J., Figueiredo P., Xavier J., Castro S.L., Brattico E. 2011 Music and emotions in the brain: familiarity matters. *PloS one* **6**, e27241.
59. Quiroga R.Q., Kreiman G., Koch C., Fried I. 2008 Sparse but not 'grandmother-cell'coding in the medial temporal lobe. *Trends in Cognitive Sciences* **12**, 87-91.
60. Sengpiel F., Stawinski P., Bonhoeffer T. 1999 Influence of experience on orientation maps in cat visual cortex. *Nature Neuroscience* **2**, 727-732.
61. Silvia P.J., Barona C.M. 2009 Do people prefer curved objects? Angularity, expertise, and aesthetic preference. *Empirical Studies of the Arts* **27**, 25-42.
62. Munar E., Gómez-Puerto G., Call J., Nadal M. 2015 Common visual preference for curved contours in humans and great apes. *PloS one* **10**, e0141106.
63. Fantz R.L. 1957 Form preferences in newly hatched chicks. *Journal of Comparative and Physiological Psychology* **50**, 422.
64. Renoult J.P., Bovet J., Raymond M. 2016 Beauty is in the efficient coding of the perceiver. *Royal Society Open Science* (DOI: 10.1098/rsos.160027).
65. Holzleitner I.J., Lee A.J., Hahn A.C., Kandrik M., Bovet J., Renoult J., Simmons D., Garrod O., DeBruine L.M., Jones B.C. 2018 A data-driven model of women's facial attractiveness reliably outperforms theory-driven models. *PsyArXiv Preprints* (DOI: 10.31234/osf.io/vhc5k).
66. Hibbard P.B., O'Hare L. 2015 Uncomfortable images produce non-sparse responses in a model of primary visual cortex. *Royal Society Open Science* **2**, 140535.
67. Pérez-Rodríguez L., Jovani R., Stevens M. 2017 Shape matters: animal colour patterns as signals of individual quality. *Proceedings of the Royal Society of London Biology B* **284**, 20162446.
68. Spehar B., Wong S., van de Klundert S., Lui J., Clifford C.W.G., Taylor R. 2015 Beauty and the beholder: the role of visual sensitivity in visual preference. *Frontiers in Human Neuroscience* **9**, 514.
69. Juricevic I., Land L., Wilkins A., Webster M.A. 2010 Visual discomfort and natural image statistics. *Perception* **39**, 884-899.
70. Redies C., Hänisch J., Blickhan M., Denzler J. 2007 Artists portray human faces with the Fourier statistics of complex natural scenes. *Network: Computational Neural Systems* **18**, 235-248.
71. Graham D.J., Field D.J. 2007 Statistical regularities of art images and natural scenes: Spectra, sparseness and nonlinearities. *Spatial Vision* **21**, 149-164.
72. Graham D.J., Redies C. 2010 Statistical regularities in art: relations with visual coding and perception. *Vision Research* **50**, 1503-1509.
73. Winkielman P., Cacioppo J.T. 2001 Mind at ease puts a smile on the face: psychophysiological evidence that processing facilitation elicits positive affect. *Journal of Personality and Social Psychology* **81**, 989.
74. Reber R., Schwarz N., Winkielman P. 2004 Processing fluency and aesthetic pleasure: is beauty in the perceiver's processing experience? *Personality and Social Psychology Review* **8**, 364-382.
75. Green N.F., Urquhart H.H., van den Berg C.P., Marshall N.J., Cheney K.L. 2018 Pattern edges improve predator learning of aposematic signals. *Behavioral Ecology* **29**, 1481-1486.

76. Landová E., Bakhshaliyeva N., Janovcová M., Pelešková Š., Suleymanova M., Polák J., Guliev A., Frynta D. 2018 Association Between Fear and Beauty Evaluation of Snakes: Cross-Cultural Findings. *Frontiers in Psychology* **9**, 333.
77. Schwarz N. 2011 Feelings-as-information theory. *Handbook of theories of social psychology* **1**, 289-308.
78. Gottlieb J., Oudeyer P.-Y., Lopes M., Baranes A. 2013 Information-seeking, curiosity, and attention: computational and neural mechanisms. *Trends in Cognitive Sciences* **17**, 585-593.
79. Fang X., Singh S., Ahluwalia R. 2007 An examination of different explanations for the mere exposure effect. *Journal of Consumer Research* **34**, 97-103.
80. Chatterjee A., Vartanian O. 2016 Neuroscience of aesthetics. *Annals of the New-York Academy of Science* **1369**, 172-194.
81. Berridge K.C., Robinson T.E., Aldridge J.W. 2009 Dissecting components of reward:'liking','wanting', and learning. *Current Opinions in Pharmacology* **9**, 65-73.
82. Kornell N. 2014 Where is the "meta" in animal metacognition? *Journal of Comparative Psychology* **128**, 143.
83. Wyer Jr R.S., Clore G.L., Isbell L.M. 1999 Affect and information processing. In *Advances in Experimental Social Psychology* (pp. 1-77), Elsevier.
84. Hubel D.H., Wiesel T.N. 1962 Receptive fields, binocular interaction and functional architecture in the cat's visual cortex. *The Journal of Physiology* **160**, 106-154.
85. Anderson D.J., Adolphs R. 2014 A framework for studying emotions across species. *Cell* **157**, 187-200.
86. Pessoa L. 2008 On the relationship between emotion and cognition. *Nature Reviews Neuroscience* **9**, 148.
87. Finucane M.L., Alhakami A., Slovic P., Johnson S.M. 2000 The affect heuristic in judgments of risks and benefits. *Journal of Behavioral Decision Making* **13**, 1-17.
88. Berridge K.C., Kringelbach M.L. 2008 Affective neuroscience of pleasure: reward in humans and animals. *Psychopharmacology* **199**, 457-480.
89. Firestone C., Scholl B.J. 2016 Cognition does not affect perception: Evaluating the evidence for" top-down" effects. *Behavioral and Brain Sciences* **39**, e229.
90. Giurfa M. 2013 Cognition with few neurons: higher-order learning in insects. *Trends in Neurosciences* **36**, 285-294.
91. Paul E.S., Mendl M.T. 2018 Animal emotion: Descriptive and prescriptive definitions and their implications for a comparative perspective. *Applied Animal Behavioral Science* **205**, 202-209.
92. Perry C.J., Baciadonna L., Chittka L. 2016 Unexpected rewards induce dopamine-dependent positive emotion–like state changes in bumblebees. *Science* **353** 1529-1531.
93. Bromberg-Martin E.S., Hikosaka O. 2009 Midbrain dopamine neurons signal preference for advance information about upcoming rewards. *Neuron* **63**, 119-126.
94. Chatterjee A. 2013 *The aesthetic brain: How we evolved to desire beauty and enjoy art*, Oxford University Press, USA.
95. Shimamura A.P., Palmer S.E. 2012 *Aesthetic Science: Connecting Minds, Brains, and Experience*. New-York, Oxford University Press.
96. Leder H., Gerger G., Brieber D., Schwarz N. 2014 What makes an art expert? Emotion and evaluation in art appreciation. *Cognition and Emotion* **28**, 1137-1147.
97. Cummings M.E., Endler J.A. 2018 25 Years of Sensory Drive: the evidence and its watery bias. *Current Zoology* **64**, 471-484.
98. Stoddard M.C., Prum R.O. 2008 Evolution of avian plumage color in a tetrahedral color space: A phylogenetic analysis of new world buntings. *The American Naturalist* **171**, 755-776.
99. Bybee S.M., Yuan F., Ramstetter M.D., Llorente-Bousquets J., Reed R.D., Osorio D., Briscoe A.D. 2012 UV photoreceptors and UV-yellow wing pigments in Heliconius butterflies allow a color signal to serve both mimicry and intraspecific communication. *The American Naturalist* **179**, 38-51.
100. Pérez-Rodríguez L., Jovani R., Mougeot F. 2013 Fractal geometry of a complex plumage trait reveals bird's quality. *Proceedings of the Royal Society of London B: Biology* **280**, 20122783.
101. Carleton K.L., Spady T.C., Streelman J.T., Kidd M.R., McFarland W.N., Loew E.R. 2008 Visual sensitivities tuned by heterochronic shifts in opsin gene expression. *BMC Biology* **6**, 22.
102. Maan M.E., Hofker K.D., van Alphen J.J.M., Seehausen O. 2006 Sensory drive in cichlid speciation. *The American Naturalist* **167**, 947-954.

103.	Thayer G.H. 1918 *Concealing-coloration in the animal kingdom: An exposition of the laws of disguise through color and pattern: being a summary of Abbott H. Thayer's discoveries*, Macmillan, New York.
104.	Poldrack R.A. 2015 Is "efficiency" a useful concept in cognitive neuroscience? *Developmental Cognitive Neuroscience* **11**, 12-17.
105.	Bullmore E., Sporns O. 2012 The economy of brain network organization. *Nature Reviews Neuroscience* **13**, 336.
106.	Hyvärinen A., Hurri J., Hoyer P.O. 2009 *Principal Components and Whitening*. London, Springer-Verlag; 93-130 pp.
107.	Jones J.P., Palmer L.A. 1987 An evaluation of the two-dimensional Gabor filter model of simple receptive fields in cat striate cortex. *Journal of Neurophysiology* **58**, 1233-1258.
108.	Olshausen B.A., Field D.J. 2004 Sparse coding of sensory inputs. *Current Opinion in Neurobiology* **14**, 481-487.
110.	Kriegeskorte N. 2015 Deep neural networks: a new framework for modeling biological vision and brain information processing. *Annual Review of Vision Science* **1**, 417-446.
111.	Tishby N., Zaslavsky N. 2015 Deep learning and the information bottleneck principle. In *Information Theory Workshop (ITW), 2015 IEEE* (pp. 1-5, IEEE.
112.	Kreutzer M., Aebischer V. 2015 The Riddle of Attractiveness: Looking for an 'Aesthetic Sense'Within the Hedonic Mind of the Beholders. In *Current Perspectives on Sexual Selection* (pp. 263-287), Springer.
113.	Mendelson T.C., Fitzpatrick C.L., Hauber M.E., Pence C.H., Rodríguez R.L., Safran R.J., Stern C.A., Stevens J.R. 2016 Cognitive phenotypes and the evolution of animal decisions. *Trends in Ecology and Evolution* **31**, 850-859.
114.	Blumstein D.T., Bouskila A. 1996 Assessment and decision making in animals: a mechanistic model underlying behavioral flexibility can prevent ambiguity. *Oikos*, 569-576.
115.	Perry C.J., Barron A.B., Chittka L. 2017 The frontiers of insect cognition. *Current Opinion in Behavioral Sciences* **16**, 111-118.
116.	Wasserman E.A., Zentall T.R. 2006 *Comparative cognition: Experimental explorations of animal intelligence*, Oxford University Press, USA.
117.	Skelhorn J., Rowe C. 2016 Cognition and the evolution of camouflage. *Proceedings of the Royal Society of London Biology B* **283**, 20152890.
118.	Cummings M.E. 2015 The mate choice mind: studying mate preference, aversion and social cognition in the female poeciliid brain. *Animal Behavior* **103**, 249-258.
119.	Rosenthal G.G. 2018 Evaluation and hedonic value in mate choice. *Current Zoology* **64**, 485-492.
120.	Schwarz N. 2017 Of fluency, beauty, and truth: Inferences from metacognitive experiences. in *Metacognitive diversity: an interdisciplinary approach*. New York, NY: Oxford University Press.
121.	Weaver K., Garcia S.M., Schwarz N., Miller D.T. 2007 Inferring the popularity of an opinion from its familiarity: A repetitive voice can sound like a chorus. *Journal of Personality and Social Psychology* **92**, 821.
122.	Loewenstein G. 1994 The psychology of curiosity: A review and reinterpretation. *Psychological Bulletin* **116**, 75.
123.	Schaeffer J.-M. 2015 *L'Expérience Esthétique*, Editions Gallimard.